\documentclass[11pt]{article}
\usepackage{times}
\usepackage{geometry}
\usepackage[utf8]{inputenc}
\usepackage{enumitem,amssymb}
\usepackage{ragged2e}
\newlist{thematic}{itemize}{8}
\setlist[thematic]{label=$\square$}
\usepackage{pifont}
\usepackage{fancyhdr}
\usepackage[USenglish]{babel}
\usepackage[utf8]{inputenc}
\usepackage[T1]{fontenc}
\usepackage{epsfig}
\usepackage{wrapfig}
\usepackage{color}

\usepackage[vflt]{floatflt}
\usepackage{longtable}
\usepackage{caption}
\usepackage{jheppub}   
\usepackage[dvipsnames,svgnames,x11names]{xcolor}
\usepackage[all]{hypcap} 
\usepackage{amssymb,amsmath}
\usepackage{amssymb}
\usepackage{amsmath, xspace}
\usepackage{graphics,graphicx} 
\usepackage{rotating}
\usepackage{multirow}
\usepackage{subfigure}
\usepackage{sectsty}
\usepackage[outercaption]{sidecap}
\sidecaptionvpos{figure}{c}

\definecolor{DarkGreen}{rgb}{0.0, 0.3, 0.0}
\definecolor{purple}{rgb}{0.5, 0.0, 0.5}
\definecolor{red}{rgb}{1, 0.0, 0.0}
\definecolor{green}{rgb}{0, 1.0, 0.0}

\def\3he{$^3{\rm He}$}

\def\lsim{\mathrel{\lower2.5pt\vbox{\lineskip=0pt\baselineskip=0pt
           \hbox{$<$}\hbox{$\sim$}}}}

\def\gsim{\mathrel{\lower2.5pt\vbox{\lineskip=0pt\baselineskip=0pt
           \hbox{$>$}\hbox{$\sim$}}}}

\begin{document}
\begin{center}
\huge
AtLAST - A five fold increase in the number of identified Strongly Lensed Galaxies in the sub-millimetre and its consequences \\  
{\normalsize This white paper was submitted to ESO Expanding Horizons in support of AtLAST}
\end{center}
\bigskip
\normalsize

\textbf{Authors:} 
Joaqu\'in Gonz\'alez-Nuevo$^{1,2}$ (gnuevo@uniovi.es); Laura Bonavera$^{1,2}$; Juan Alberto Cano$^{1,2}$; David Crespo$^{1,2}$; Rebeca Fern\'andez-Fern\'andez$^{1,2}$; Valentina Franco$^{1,2}$; Marcos M. Cueli$^{1,2}$; Jos\'e Manuel Casas$^{1,2}$; Tom J. L. C. Bakx$^3$\\
$^1$Departamento de Fisica, Universidad de Oviedo, C. Federico Garcia Lorca 18, 33007 Oviedo, Spain.\\
$^2$Instituto Universitario de Ciencias y Tecnologías Espaciales de Asturias (ICTEA), C. Independencia 13, 33004 Oviedo, Spain.\\
$^3$Department of Space, Earth, \& Environment, Chalmers University of Technology, Chalmersplatsen 4, Gothenburg SE-412 96, Sweden\\


\textbf{Science Keywords:} 
cosmology: large-scale structure of universe, cosmology: dark energy, cosmology: dark matter, cosmology: lenses, cosmology: observations

 \captionsetup{labelformat=empty}
\begin{figure}[h]
   \centering
   \captionsetup{width=0.9\textwidth}
\includegraphics[width=1\textwidth]{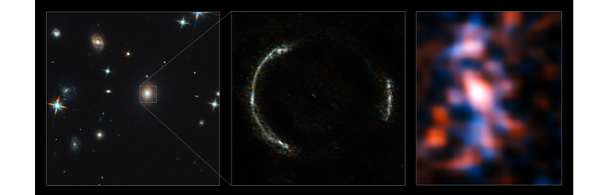}
   \caption{Three views of the strongly lensed galaxy SDP.81: foreground lens and faint Einstein ring (left), sharp ALMA image of the ring (centre), and lens‑model reconstruction (right) revealing multiple cold dust clouds where stars and planets are forming. Credit:
ALMA (NRAO/ESO/NAOJ)/Y. Tamura (The University of Tokyo)/Mark Swinbank (Durham University)}
\end{figure}
\vspace{-15mm}

\setcounter{figure}{0}
\captionsetup{labelformat=default}


\pagebreak
\section*{Abstract}
Strong gravitational lensing is a powerful probe of cosmology, dark matter (DM), and high-redshift galaxy evolution, but current samples of strongly lensed galaxies (SLGs) remain far too small to exploit its full potential. \textit{Herschel}’s submillimeter (submm) surveys demonstrated that submm selection provides the most efficient and least biased route to identifying high-redshift SLGs, yet produced only a few hundred systems over limited, heterogeneous fields. Achieving the thousands of SLGs required for precision cosmology and detailed studies of distant dusty star-forming galaxies demands a new, wide-area, homogeneous sub-mm survey. A facility like AtLAST, capable of extending \textit{Herschel}-like methodologies to much larger areas, is uniquely positioned to deliver the order-of-magnitude increase in SLG numbers needed for transformative progress.

\section{Scientific context and motivation}
Strong gravitational lensing is a well-known phenomenon caused by the deflection of light from a distant background source by the gravitational field of a massive foreground object, such as a galaxy, group, or cluster. When the source and lens are well aligned with the observer, this deflection produces multiple images, arcs, or rings of the background source (Schneider et al., 1992).

Before wide-area sub-mm surveys, the discovery of SLGs relied on optical, near-infrared, radio, and millimeter techniques. Optical spectroscopic searches like SLACS identified lenses via two redshifts—absorption lines from a foreground early-type galaxy and emission lines from a background source—producing 85 systems but limited to $ z\leq 0.7$ (Auger et al., 2009). Quasar-based searches such as SQLS (Oguri et al., 2006) or radio surveys like CLASS (Browne et al., 2003) used large parent samples ($\sim50,000$ quasars for SQLS and $\sim9,000$ radio sources for CLASS) but achieved very low efficiencies ($\sim0.2\%$ in CLASS), yielding only tens of confirmed lenses. Wide-field imaging surveys such as DES extended the area (>5000 deg$^2$) and used machine learning to identify arc-like morphologies around luminous red galaxies, yet still obtained modest candidate densities ($\simeq0.078$ deg$^ {-2}$; Rojas et al., 2022). Millimeter surveys like those from the South Pole Telescope (SPT) used bright mm flux densities similar to sub-mm selection but, with smaller fields, reached densities of $\simeq0.15-0.3$ deg$^ {-2}$, producing only a few dozen candidates (Spilker et al., 2016). Overall, optical and radio methods yielded small samples (tens to low hundreds) with low number densities and strong biases toward low redshift or specific source populations.

The \textit{Herschel} Space Observatory revolutionized SLG identification by exploiting the steep number counts of high-redshift dusty star-forming galaxies (DSFGs) at sub-mm wavelengths. A simple flux cut at $500 \mu m$ ($S_{500} \geq 100$ mJy), combined with removal of local galaxies and blazars, achieved near-100\% efficiency in isolating SLGs (Negrello et al., 2010). This approach yielded surface densities around 0.3 deg$^ {-2}$—far higher than optical or radio surveys—and produced the first statistically significant sub-mm-selected SLG samples ($\simeq80$ in H-ATLAS, Negrello et al., 2017; $\simeq77$ in HerMES, Nayyeri et al., 2016). Extending to fainter flux cuts (80–100 mJy) or using statistical/multi-wavelength methods (HALOS, González-Nuevo et al., 2012; SHALOS, 2019; FLASH, Bakx et al., 2024) increased densities to 1.5–2 deg$^ {-2}$, identifying hundreds of SLGs across \textit{Herschel}’s $\simeq 600-700$ deg$^2$ coverage. \textit{Herschel} thus provided the largest homogeneous SLG samples and a simple, efficient, and minimally contaminated selection function ideal for lensing statistics.

Upcoming wide-area surveys will greatly expand SLG samples using complementary techniques. \href{https://www.cosmos.esa.int/web/euclid}{\textit{\underline{Euclid}}} will deliver diffraction-limited optical/NIR imaging across $\simeq15,000$ deg$^2$, enabling identification of $\simeq 10^5$ galaxy–galaxy strong lenses ($\simeq6.7$ deg$^ {-2}$; Laureijs et al., 2011). Additional NISP data will reveal highly magnified H$_\alpha$ emitters and lensed NIR sources, adding several tens of thousands. The Square Kilometre Array (\href{https://www.skao.int/en}{\underline{SKA}}) will exploit magnification bias at the bright end of the HI mass function, potentially identifying $\simeq10^5$ lensed HI sources ($\simeq12-13$ deg$^{-2}$; Serjeant, 2014). These facilities will yield unprecedented samples, but their selection functions, wavelength regimes, and redshift sensitivities differ from sub-mm–selected DSFG lenses, making them complementary, not replacements.

Despite the promise of \textit{Euclid} and SKA, sub-mm selection remains uniquely powerful: it preferentially identifies high-redshift DSFGs ($z \simeq 2-4$), is minimally biased by lens properties, suffers negligible contamination, and directly probes dust-obscured star formation at the peak epoch of galaxy assembly.

\vspace{-3mm}
\section{Science case}
Strong gravitational lensing is one of the most powerful natural tools in observational astrophysics, allowing us to probe the mass distribution of galaxies and their DM halos, to magnify intrinsically faint high-redshift galaxies, and to obtain spatial resolutions far beyond the capabilities of current instrumentation. Lensing provides a purely gravitational measurement of mass, independent of stellar or gas tracers, and the magnification produced by lensing acts as a “cosmic telescope” that reveals the internal structure of distant galaxies at sub-kiloparsec scales. As such, samples of strong lenses are uniquely suited to address a range of forefront questions in cosmology, galaxy evolution, and DM physics (Vegetti et al., 2024).

Large samples of SLGs, reaching into the thousands, would deliver a transformative advance for cosmology. Strong-lensing statistics are sensitive to the abundance and evolution of DM halos, the geometry of the Universe, and the properties of dark energy (DE). With $\sim 100$ lenses one can already constrain $\Omega_\Lambda$ at the few-percent level, while sample sizes approaching $\sim 1000$ would allow constraints on the DE equation-of-state parameter $w$ comparable to those from CMB experiments (Eales, 2015). Additionally, high-redshift lensed sources identified through sub-mm selection probe a unique redshift regime ($z \simeq 2-4$), dramatically increasing the sensitivity of strong-lensing cosmological tests and allowing measurements that cannot be achieved using the low-redshift lenses typically found in optical surveys (Gonz\'{a}lez-Nuevo et al., 2012).

Large SLG samples also offer an exceptional laboratory for studying the nature of DM and the evolution of DM halos. Strong lensing is sensitive to the abundance of low-mass subhalos (down to $\sim 10^8 - 10^9 M_\odot$), enabling stringent tests of the cold DM paradigm at small scales (Negrello et al., 2017). The mass profiles of individual lenses can be reconstructed with high precision, revealing how halo structure evolves with redshift and allowing comparisons with theoretical predictions (Wardlow et al., 2012). Current samples already show tension with some standard halo-evolution models, suggesting that significantly larger samples are needed to map the statistics of DM halos across cosmic time and to uncover potential departures from $\Lambda$CDM predictions.

Beyond cosmology and DM, thousands of strongly lensed DSFGs would provide an unprecedented window onto galaxy evolution during the peak of cosmic star formation. The magnification produced by lensing boosts faint, otherwise undetectable DSFGs above detection limits, enabling detailed studies of their star formation rates, dust content, gas dynamics, and morphology (Gonz\'{a}lez-Nuevo et al., 2012). Approximately 65\% of bright sub-mm selected lenses have intrinsic flux densities below Herschel’s confusion limit, demonstrating that lensing uniquely grants access to the bulk of the high-redshift star-forming galaxy population. With spatial resolutions in the source plane reaching $\sim 0.01''$, large samples of SLGs allow systematic exploration of the internal structure and dynamical processes in thousands of galaxies at epochs currently inaccessible to direct imaging (Bakx et al., 2018).

Despite the progress made with \textit{Herschel} and other surveys, current samples remain too small, too heterogeneous, and too shallow to allow these scientific goals to be fully realised. \textit{Herschel}-based methods have produced only a few hundred confirmed SLGs across a patchwork of fields, with typical densities of 0.1–2 deg$^{-2}$ depending on the selection method. Optical and radio surveys have added only tens to low hundreds of additional systems, often at lower redshift and with complex or poorly understood selection functions. This limited census lacks the statistical power required for high-precision cosmology, for robust mapping of halo substructure, or for detailed population-level studies of high-redshift DSFGs. Moreover, the available data were not acquired through a single homogeneous survey, making it difficult to control selection biases or to build uniform samples on the scales necessary for next-generation lensing science.

Sub-mm selection remains uniquely efficient and physically well-motivated for identifying strongly lensed high-redshift galaxies. The combination of steep DSFG number counts, strong magnification bias, minimal contamination from the lens, and negligible dust-attenuation effects makes sub-mm surveys by far the cleanest and most complete method for assembling large samples of SLGs at $z \simeq 2-4$. While \textit{Herschel} provided the proof of concept, only a next-generation facility like AtLAST, with its wide instantaneous field of view, high mapping speed, and ability to cover 2000-3000 deg$^2$ uniformly, can extend this methodology to the areas required to identify $\sim 5000$ SLGs. A large, homogeneous sub-mm survey is therefore essential: it is the only feasible pathway to reaching the order-of-magnitude increase in SLG numbers needed to unlock the full scientific potential of strong gravitational lensing in cosmology, DM, and galaxy evolution.

\section{Technical requirements}
No existing or planned facility can provide the combination of sky coverage, angular resolution, and submm sensitivity required to expand the census of strongly lensed galaxies from a few hundred today to the many thousands needed for precision cosmology, DM studies, and resolved investigations of high-redshift dusty star-forming galaxies. Achieving these goals demands a telescope capable of conducting wide, homogeneous sub-mm surveys while maintaining confusion-limited sensitivity and the ability to detect faint, highly magnified DSFGs over large sky areas. AtLAST uniquely meets these requirements (van Kampen et al., 2025). Its 50-m aperture delivers $\sim 1.5''$ diffraction-limited resolution at 950 GHz—crucial for suppressing confusion noise, resolving blended Herschel sources, and enabling accurate identification of strongly lensed systems selected via magnification bias. AtLAST’s instantaneous two-degree field of view (Gallardo et al., 2024) and mapping speeds orders of magnitude faster than ALMA make it the only facility capable of surveying thousands of square degrees uniformly in the submm. 

Next-generation instrumentation on AtLAST will include large-format cameras with over one million pixels, delivering unprecedented sensitivity and dramatically increasing the surface density of high-redshift submm galaxies. This will boost the signal-to-noise ratio of magnification-bias measurements by orders of magnitude. Its broad frequency coverage (30–950 GHz; Mroczkowski et al., 2025) enables accurate colour selection and photometric redshifts for background sources, while simultaneous spectroscopic and continuum observations (Gallardo et al., 2024) will allow detailed characterization of lens and source populations. These combined capabilities represent a transformative leap beyond current facilities, enabling science that cannot be achieved by \href{https://almascience.eso.org}{\underline{ALMA}}, \href{https://science.nasa.gov/mission/webb/}{\underline{JWST}}, \href{https://www.cosmos.esa.int/web/euclid}{\textit{\underline{Euclid}}}, or \href{https://rubinobservatory.org}{\textit{\underline{Rubin}}}. Last but not least, AtLAST is pioneering sustainable astronomy: it is the first astronomical facility that, since its initial design phase, has been researching engineering solutions for a tailored, off-grid renewable energy system inclusive of a hybrid battery/hydrogen storage that can supply 100\% of the power needed for day and night-time telescope operations (Viole et al., 2023; Kiselev et al., 2024).

\bigskip
\noindent\textbf{References:} \small{Auger, M.W, et al., 2009, ApJ, 705, 1099  $\bullet$ Bakx T.~J.~L.~C., et al., 2024, MNRAS, 527, 8865 $\bullet$ Bakx T.~J.~L.~C., et al., 2018, MNRAS, 473, 1751 $\bullet$ Browne, I.W.A., et al., 2003, MNRAS, 341, 13 $\bullet$ Eales, S.A., 2015, MNRAS, 446, 3224 $\bullet$ Gallardo, P.A., et al., 2024, Proc. of the SPIE, 13094, 1309428 $\bullet$ Gonz{\'a}lez-Nuevo J., et al., 2012, ApJ, 749, 65 $\bullet$ Gonz{\'a}lez-Nuevo J., et al., 2019, A\&A, 627, A31 $\bullet$ van Kampen, E., et al., 2025, Open Research Europe, 4, 122 $\bullet$ Kiselev, A., et al., 2024, Proc. of the SPIE, 13094, 130940E $\bullet$ Laureijs, R., et al., 2011, arXiv:1110.3193 $\bullet$ Mroczkowski, T., et al., 2025, A\&A, 694, A142 $\bullet$ Negrello, M. et al., 2010, Science, 330, 6005 $\bullet$ Negrello, M., et al., 2017, MNRAS, 465, 3558 $\bullet$ Oguri, M., et al. 2006, ApJ 132, 999 $\bullet$ Rojas, K., et al., 2022, A\&A, 668, A73 $\bullet$ Serjeant, S., 2014, ApJL, 793, L10 $\bullet$ Schneider, P., Ehlers, J., \& Falco, E. E. 1992, Gravitational Lenses $\bullet$ Spilker, J.S., et al., 2016, ApJ, 826, 112 $\bullet$ Vegetti, S., et al., 2024, Space Science Reviews, 220, 58 $\bullet$ Viole, I., et al., 2023, Energy, 282, 128570 $\bullet$ Wardlow, J.L., et al., 2012, ApJ, 762, 59} 

\end{document}